\documentclass[aps,twocolumn,prd,superscriptaddress,nofootinbib,floatfix]{revtex4-2}

\usepackage{tikz}
\usetikzlibrary{arrows.meta}

\usepackage{mathtools}
\usepackage{amsfonts}
\usepackage{amssymb}
\usepackage{mathrsfs}
\usepackage{bbm}
\usepackage{slashed}
\usepackage{amsmath}
\usepackage{bm}
\usepackage{bbold}

\usepackage{graphicx}
\usepackage{color}
\usepackage{colortbl}
\usepackage{array}

\usepackage{float}
\usepackage{placeins}
\usepackage{booktabs}
\usepackage[caption=false]{subfig}
\usepackage{makecell}
\usepackage{tabstackengine}

\usepackage{xspace}
\usepackage{hyperref}
\usepackage[nameinlink]{cleveref}
\usepackage{bookmark}
\usepackage{siunitx}

\usepackage{xifthen}
\usepackage{xcolor}
\hypersetup{
	colorlinks,
	linkcolor={red!75!black},
	citecolor={blue!75!black},
	urlcolor={blue!75!black}
}

\usepackage[utf8]{inputenc}
\allowdisplaybreaks[4]
\usepackage[normalem]{ulem}

\usepackage[draft,authormarkup=none]{changes}
\definechangesauthor[name={Yi-zhen Huang}, color=red]{YH}

\newcommand{\MeV}{\;\text{MeV}}
\newcommand{\GeV}{\;\text{GeV}}

\newcommand{\feyn}[1]{
	\setbox0=\hbox{\ensuremath{#1}}
	\hbox to\wd0{\hbox to0pt{\hbox to\wd0{\hss/\hss}\hss}\box0}}

\def\Eq#1{Eq.~\labelcref{#1}}

\def\Fig#1{Fig.~\labelcref{#1}}
\def\Tab#1{Tab.~\labelcref{#1}}
\def\sec#1{Sec.~\labelcref{#1}}
\def\app#1{App.~\labelcref{#1}}

\setkeys{Gin}{width=0.48\textwidth}
\graphicspath{{./figures/}}

\newcommand{\getLanzhouAffiliation}{\affiliation{School of Physical Science and Technology, Lanzhou University, Lanzhou 730000, China}}

\newcommand{\getDalianAffiliation}{\affiliation{School of Physics, Dalian University of Technology, Dalian, 116024, P.R. China}}

\newcommand{\getGiessenAffiliation}{\affiliation{Institut f\"ur Theoretische Physik, Justus-Liebig-Universit\"at Gie\ss en, 35392 Gie\ss en, Germany}}

\newcommand{\getEnergyAffiliation}{\affiliation{Energy Singularity Fusion Power Technology (SH) Ltd., Shanghai, China}}

\begin{document}
\raggedbottom
\title{Functional renormalization group study of the jet quenching parameter near the QCD critical end point}
	
\author{Yi-zhen Huang}
\getDalianAffiliation

\author{Shi Yin}
\getGiessenAffiliation

\author{Sheng-nan Han}
\getDalianAffiliation
\getLanzhouAffiliation

\author{Jing Wu}
\email{wujing@mail.dlut.edu.cn}
\getDalianAffiliation

\author{Feng Li}
\email{lifengphysics@gmail.com}
\getEnergyAffiliation

\author{Wei-jie Fu}
\getDalianAffiliation

\begin{abstract}

We investigate the jet quenching parameter $\hat{q}$ in the QCD phase diagram within a QCD-assisted low-energy effective theory using the functional renormalization group (fRG). Following the formalism that relates $\hat{q}$ to the spectral functions of the chiral order-parameter field, we compute the $\sigma$ and $\pi$ meson contributions to $\hat{q}$ at finite temperature and baryon chemical potential from analytically continued mesonic two-point functions. We find that $\hat{q}$ receives appreciable contributions mainly above the chiral phase boundary and exhibits a pronounced enhancement at large baryon chemical potential as the chiral crossover sharpens toward the critical end point (CEP), a behavior consistent with the picture of partonic critical opalescence (PCO) -- a pronounced enhancement of jet transverse momentum broadening induced by the critical $\sigma$ field fluctuations.

\end{abstract}

\maketitle

\section{Introduction}
\label{sec:intro}

Whether the QCD phase diagram possesses a critical end point (CEP), i.e., the end point of a first-order phase transition line in the regime of large baryon chemical potential, and if so, where it is located, remains one of the most fundamental open questions in strong interaction physics. At small baryon chemical potential, it is found that the transition from the quark-gluon plasma (QGP) to hadronic matter is a smooth crossover~\cite{Aoki:2006we, Bazavov:2011nk}, while at large baryon chemical potential, the existence and location of a CEP are still under active investigations~\cite{Fu:2019hdw, Gao:2020fbl, Gunkel:2021oya, Pawlowski:2025jpg, Fu:2026qnl}, see also \cite{Stephanov:2004wx, Fukushima:2013rx, Dupuis:2020fhh, Fu:2022gou, Fischer:2026uni} for relevant reviews. A widely adopted strategy for probing QCD criticality is through the higher-order cumulants of conserved charges~\cite{Stephanov:2008qz, Stephanov:2011pb, Bzdak:2019pkr, STAR:2020tga, STAR:2025zdq} such as the baryon number~\cite{Fu:2015amv, Fu:2016tey, Fu:2021oaw, Fu:2023lcm, Lu:2025cls, Lu:2026ezr}. In the heavy-ion collision experiments, the beam energy scan program of second phase (BES-II) at RHIC has delivered high-statistics measurements of net-proton fluctuations~\cite{STAR:2025zdq}, revealing suggestive deviations from non-critical baselines at certain collision energies. Specifically, recently it is predicted from theoretical calculations that there is a peak structure in the kurtosis of baryon number fluctuations as a function of the collision energy in the fixed-target energy range~\cite{Fu:2023lcm, Lu:2026ezr}. In addition, the temperature fluctuations~\cite{Chen:2025vwl}, spin fluctuations and correlations~\cite{Chen:2024hki},  light nuclei yield ratios~\cite{Sun:2020zxy, Han:2025sld} have also been proposed as possible complementary probes of the CEP. 

The probes mentioned above are all formed in the late stage of the collision, around the chemical and kinetic freeze-outs, when the system has already cooled down close to the hadronization temperature. As a consequence, they primarily reflect the state of the medium at these freeze-out surfaces, which lie near or below the chiral phase boundary in the QCD phase diagram. Even though the freeze-out curves pass relatively close to the CEP at large baryon chemical potential, they do not necessarily coincide with each other, so the critical signal imprinted on the medium near the CEP has to survive from a non-trivial evolution through hadronization and subsequent hadronic rescatterings. The extent to which these soft probes can faithfully reflect the direct information about the CEP is therefore not clear. This motivates us to search for observables that are produced at the earlier stage of the collision and sensitive to the medium throughout its entire evolution. Jets are the natural candidates: originating from hard scatterings in the initial impact, they traverse the QGP while it expands and cools, and lose energy through interactions with the local medium. The resulting transverse momentum broadening is characterized by the jet quenching parameter $\hat{q}$~\cite{Baier:2002tc, Majumder:2008zg}, which directly probes the local transport properties of the medium. 
Experimentally, a hint along this direction has already been seen: the nuclear modification factor $R_{AA}$ for hard hadrons measured in the BES program shows an intriguing additional suppression at certain centralities in Au+Au collisions at $\sqrt{s_{\rm NN}} = 14.5\GeV$~\cite{STAR:2017ieb}, an observation whose theoretical implications remain to be fully explored. In a recent study, Wu {\it et al.} computed $\hat{q}$ within the quark-meson model in a framework that relates $\hat{q}$ to the spectral function of the chiral order parameter. They obtained this spectral function by mapping to the Ising model and identified a pronounced enhancement of $\hat{q}$ near the CEP, which they interpreted as the partonic critical opalescence (PCO), in analogy to the enhancement of light scattering near a classical liquid-gas critical point. The critical behavior was thus captured through an external universality-class input: as also shown in their work, no such enhancement appears when the chiral order-parameter spectral function is instead obtained from a one-loop perturbative calculation.

%
\begin{figure*}[t]
\centering
\includegraphics[width=0.8\textwidth]{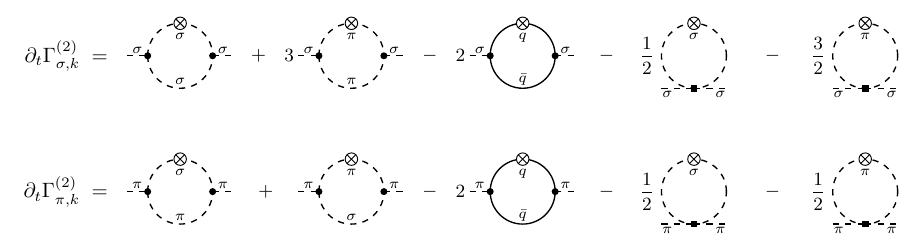}
\caption{Diagrammatic representation of the flow equations for the mesonic two-point correlation functions. The three- and four-point vertices are represented by the full circles and squares, respectively. The crossed circles denote the infrared regulators.}
\label{fig:Gamma2_flow}
\end{figure*}
%

Following the formalism developed in~\cite{Wu:2022vbu}, in this work we compute $\hat{q}$ at finite temperature and baryon chemical potential within a QCD-assisted low-energy effective theory based on the Polyakov-loop quark-meson (PQM) model~\cite{Fu:2021oaw, Fu:2023lcm} and the functional renormalization group (fRG)~\cite{Wetterich:1992yh, Pawlowski:2005xe, Dupuis:2020fhh, Fu:2022gou}. The fRG provides a non-perturbative framework that systematically incorporates quantum and thermal fluctuations across momentum scales, and is capable of capturing the critical behavior in the vicinity of the CEP. For recent progress in the studies of QCD in the vacuum see e.g., \cite{Mitter:2014wpa, Braun:2014ata, Cyrol:2016tym, Cyrol:2017ewj, Corell:2018yil, Fu:2022uow, Fu:2024ysj, Fu:2025hcm, Ihssen:2024miv, Pawlowski:2025jpg, Zhang:2025ofc, Cui:2026bod} and at finite temperature and densities see e.g., \cite{Fu:2019hdw, Braun:2020ada,  Fu:2021oaw, Fu:2023lcm, Fu:2024rto, Tan:2024fuq, Pawlowski:2025jpg, Tan:2025bsv, Fu:2026qnl}.

In our QCD-assisted setup, the running Yukawa coupling between quarks and mesonic fields is determined by matching its RG flow to that obtained in the first-principles fRG calculation of 2+1-flavor QCD at finite temperature~\cite{Fu:2019hdw}. This strategy has been successfully employed in recent studies of baryon number fluctuations and critical dynamics near the CEP~\cite{Fu:2023lcm}. Our results are consistent with the PCO scenario: $\hat{q}$ exhibits a pronounced enhancement at large baryon chemical potential as the chiral crossover sharpens toward the CEP, suggesting that jet observables may serve as a valuable tool for probing the QCD phase structure.

\section{Theoretical basis}
\label{sec:jet} 

%
\begin{figure*}[t]
\centering
\includegraphics[width=0.8\textwidth]{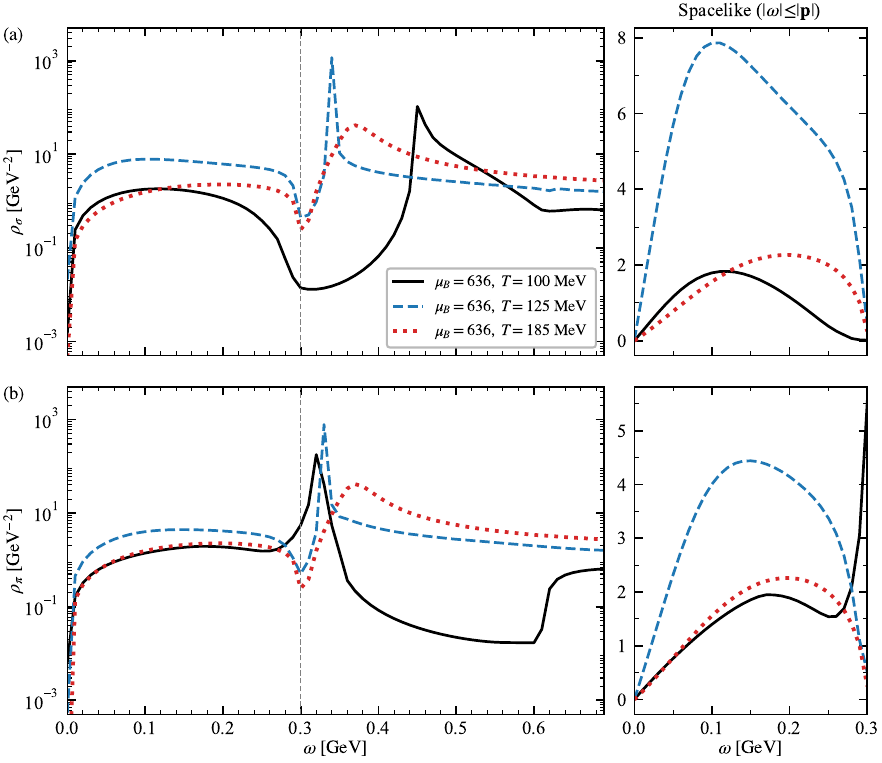}
\caption{The $\sigma$ (a) and $\pi$ (b) spectral functions at $\mu_B = 636\MeV$ and $|\bm{p}| = 0.3\GeV$ for three temperatures: $T = 100\MeV$ (black solid), $T = 125\MeV$ (blue dashed), and $T = 185\MeV$ (red dotted). Left panels show the full spectral function on a logarithmic scale, with the vertical dashed line indicating the light cone $\omega = |\bm{p}|$. Right panels show the spacelike region $|\omega| \leq |\bm{p}|$ on a linear scale.}
\label{fig:spectral}
\end{figure*}
%

\subsection{Jet quenching parameter}

The jet quenching parameter is defined as
\begin{align}
    \hat{q} \equiv \frac{\langle \bm{p}_{\perp}^{\,2} \rangle}{t}= \sum_{p} \bm{p}_{\perp}^{\,2}\,\frac{W(p)}{t}\,.\label{eq:qhatdefine}
\end{align}
Here, $t$ denotes the time during which the jet quark propagates through the medium, $p$ is the momentum exchanged between the jet quark and the medium, $\bm{p}_{\perp}$ is the component of $\bm{p}$ transverse to the initial jet momentum, and $W(p)$ is the squared matrix element averaged over color, isospin, and spin degrees of freedom, which is given by~\cite{Majumder:2012sh} 
\begin{align}
    W(p)=\frac{1}{2N_c N_f}\Big|\Big\langle q\!+\!p;X\Big|\mathcal{T}\,e^{-i\!\int_{0}^{t}dt'\,H_{\mathrm{I}}(t')}\Big|q;M\Big\rangle\Big|^2\,,\label{eq:matrixW}
\end{align}
where $q$ denotes the momentum of the jet quark before propagating through the medium. $\lvert q;M\rangle$ and $\lvert q+p;X\rangle$ represent the initial and final states of the combined jet–medium system respectively, where $M$ and $X$ denote the medium. The operator $\mathcal{T}$ stands for the time-ordering operator and $H_{\mathrm{I}}(t)$ the interaction Hamiltonian in the interaction picture. 

In this work, we describe the jet–medium interaction within the two-flavor PQM model, in which the interaction Hamiltonian is taken to be
\begin{align}
    H_{\mathrm{I}} = g_{\mathrm{eff}} \int d^{3}x\, \bar{q}(x)\big(\sigma(x) + i\gamma_{5}\,\bm \tau\cdot\bm \pi(x)\big)\,q(x)\,. \label{eq:HI}
\end{align}
Here $\bm \tau$ denotes the Pauli matrices in isospin space, $g_{\mathrm{eff}}$ denotes the effective quark-meson Yukawa coupling entering the jet-medium interaction. In the present framework, it is identified with the renormalized Yukawa coupling introduced below, namely $g_{\mathrm{eff}}=\bar h_{k=0}/2$.

In evaluating Eq.~(\ref{eq:matrixW}), we treat the jet quark as a free particle state and factorize the full initial and final states as 
\begin{align}
    \hspace{-0.1cm}\lvert q;M\rangle = \lvert q\rangle \otimes \lvert M\rangle \;\; \text{and} \;\; \lvert q+p;X\rangle = \lvert q+p\rangle \otimes \lvert X\rangle\,,
\end{align}
respectively. The free quark states are normalized in a finite volume $V$ as 
\begin{align}
    q(x)\,\lvert q\rangle = \frac{1}{\sqrt{2 E_q V}} e^{i q\cdot x} u(q)\,,
\end{align}
where $q(x)$ stands for the quark field and $u(q)$ the Dirac spinor. Expanding the time-ordered exponential to leading order in $H_{\mathrm{I}}$, one can obtain the explicit form of the jet quenching parameter $\hat{q}=\hat{q}_{\sigma}+\hat{q}_{\pi}$ with
\begin{align}
    &\quad\hat{q}_{\sigma/\pi}\nonumber\\[2ex]
    &=\frac{g_{\mathrm{eff}}^2 }{(2\pi)^3 N_c N_f}d_{\sigma/\pi}\int d^{3}\bm p \frac{\bm p^2_{\perp}}{ E_q E_{q+p}}q\cdot(q+p)\widetilde{G}_{\sigma / \pi}(p)\,,\label{eq:qhatSeparate}
\end{align}
where
\begin{align}
    \widetilde{G}_{\sigma}(p) &= \int d^{4}x\, \langle M|\sigma(0)\sigma(x)|M \rangle e^{ip\cdot x}\,, \label{eq:G-sigma}\\
    \widetilde{G}_{\pi}(p) \delta_{ij}&= \int d^{4}x\, \langle M|\pi_i(0) \pi_j(x)|M \rangle e^{ip\cdot x}\,,\label{eq:G-pi}
\end{align}
represent the correlation functions for the $\sigma$ and $\pi$ fields respectively; $d_{\sigma}=1$ and $d_{\pi}=3$ denote their respective degeneracy. In deriving the expression above, the discrete sum in \Eq{eq:qhatdefine} is replaced by integral because we are dealing with a continuum system. We further impose the on-shell conditions $q^{2} = (q+p)^{2} = 0$ for the initial and final jet quarks, which restricts the energy transfer to $|p_0| < |\bm{p}|$, see \cite{Wu:2022vbu} for more details, so that only the spacelike region of the spectral function contributes to the $\hat{q}$.

Using the Kubo-Martin-Schwinger (KMS) relation, the correlation functions in \Eq{eq:G-sigma} and \Eq{eq:G-pi} can be expressed in terms of the spectral functions $\rho_{\sigma/\pi}$ as~\cite{PhysRev.115.1342}
\begin{align}
    \widetilde{G}_{\sigma/\pi}(p) =\frac{1}{ e^{\beta p_0}-1}{\rho}_{\sigma/\pi}(p)\,,
\end{align} 
with $\beta=1/T$ and the temperature $T$.

\subsection{Spectral functions}

In this work, the spectral functions $\rho_{\sigma/\pi}$ are obtained from the retarded mesonic two-point correlation functions, whose scale dependence in fRG is determined by the Wetterich flow equation for the effective action $\Gamma_k$,
\begin{align}
    \partial_t\Gamma_k=\frac{1}{2}\mathrm{STr}[\partial_t R_k(\Gamma_k^{(2)}+R_k)^{-1}]\,,
\end{align}
with 
\begin{align}
    \Gamma_k^{(2)}[\Phi]=\frac{\delta^2 \Gamma_k[\Phi]}{\delta \Phi^2}\,,
\end{align}
where $t=\ln(k/\Lambda)$ is the RG time, and $\Lambda$ denotes some ultraviolet cutoff. Here, $R_k$ is the infrared regulator, which suppresses fluctuations of momenta $p \lesssim k$. Its explicit expressions are given in \Eq{eq:RkB} and \Eq{eq:RkF}. By taking the second functional derivative of both sides with respect to the mesonic fields, one obtains the flow equations for the two-point correlation functions, which are represented diagrammatically in \Fig{fig:Gamma2_flow}. In this work, we adopt the Euclidean scale-dependent effective action for the two-flavor PQM model as follows,
\begin{align}
    \Gamma_k[\Phi]&= \int_{0}^{1/T} d\tau \int d^3\bm{x}\, \Big[Z_{q,k}\bar{q}\Big(\gamma_\mu \partial_\mu - \gamma_0(\hat{\mu}+igA_0)\Big)q \nonumber\\
    &\quad + \frac{1}{2} Z_{\phi,k}\big(\partial_\mu \phi\big)^2 + h_k\bar{q}\big(T^0\sigma + i\gamma_5 \bm{T}\cdot\bm{\pi}\big)q \nonumber\\
    &\quad + V_k(\rho, A_0) - c\sigma \Big]\,,\label{eq:action}
\end{align}
with the fields $\Phi=(q, \bar q, \phi)$, i.e., the $u$, $d$ two-flavor quark field $q=(q_u, q_d)^{\top}$ and the sigma and pion meson fields $\phi=(\sigma, \bm\pi)$. The subscript $k$ represents the renormalization group (RG) scale. The generators of $U(2)$ group in the flavor space read $(T^0,\bm T)=1/2 (\mathbb{1}, \bm \tau)$ with the Pauli matrices $\bm \tau$. The quarks interact with the mesonic fields through the Yukawa coupling $h_k$. The effective potential $V_k(\rho, A_0)$ is comprised of a pure glue part and a matter part,
\begin{align}
    V_k(\rho, A_0) = V_{\mathrm{glue},k}(A_0) + V_{\mathrm{mat},k}(\rho, A_0)\,,\label{eq:V_decomp}
\end{align}
where $V_{\mathrm{glue},k}(A_0)$ denotes the glue potential, i.e., the Polyakov-loop potential, which is used from \cite{Fu:2021oaw}. The $O(4)$-invariant matter potential $V_{\mathrm{mat},k}(\rho, A_0)$ with $\rho = \phi^2/2$ is computed directly via its flow equation. Their explicit forms and the determination of model parameters are given 
in \app{app:setup}. The last term with the strength $c$ breaks the chiral symmetry explicitly. Here $Z_{q,k}$ and $Z_{\phi,k}$ denote the wave functions of the quark and meson fields, respectively. The quark chemical potential matrix in flavor space reads $\hat{\mu}=\mathrm{diag}(\mu,\mu)$ with $\mu=\mu_B/3$, where $\mu_B$ is the baryon chemical potential. The temporal component of the gluon background field $A_0$ couples to the quarks through the gauge coupling $g$, encoding the effects of confinement via the Polyakov loop.

The retarded two-point correlation function is obtained from the imaginary-time two-point correlation function through an analytic continuation
\begin{align}
    \Gamma_{k,\phi}^{(2),R}(\omega,\bm p)=-\lim\limits_{\epsilon \to 0}  \Gamma_{k,\phi}^{(2),E}(-\mathrm{i}(\omega+\mathrm{i}\epsilon),\bm p).
 \end{align}
Note that the real-time information is not reconstructed a posteriori from discrete Euclidean data, but is obtained by analytically continuing the flow equations themselves. The use of 3D regulators makes the Matsubara sums analytically tractable and leads to explicit threshold functions, on which the continuation in the external frequency can be performed explicitly~\cite{Tripolt:2014wra, Jung:2016yxl, Fu:2024rto}. As a result, the retarded two-point functions are computed directly within the fRG flow. The retarded propagator is related to the two-point correlation function as follows,
\begin{align}
    (G^R_{\phi}(p))^{-1}=\Gamma_{k=0,\phi}^{(2),R}(p),
\end{align}
The spectral functions are given by the imaginary part of the retarded propagator, viz.
\begin{align}
    {\rho}_{\sigma/\pi}(p)=-2\mathrm{Im}G^R_{\sigma/\pi}(p).
\end{align}

\section{Numerical results}
\label{sec:results}

%
\begin{figure}[t]
\centering
\includegraphics[width=\columnwidth]{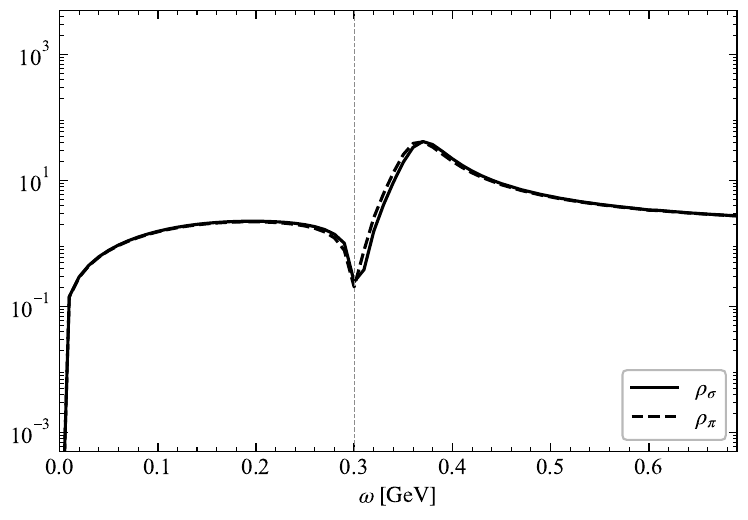}
\caption{Comparison of the $\sigma$ (solid) and $\pi$ (dashed) spectral functions at $\mu_B = 636\MeV$, $T = 185\MeV$, and $|\bm{p}| = 0.3\GeV$. The near-degeneracy of the two spectral functions is consistent with the approximate restoration of chiral symmetry at this temperature.}
\label{fig:degeneracy}
\end{figure}
%

%
\begin{figure*}[t]
\centering
\includegraphics[width=1.0\textwidth]{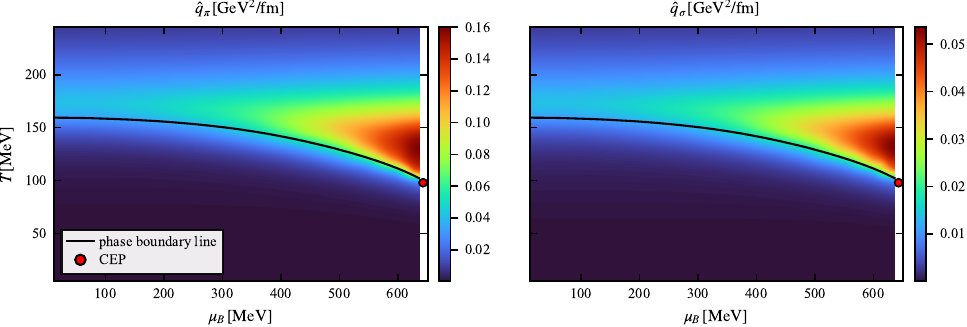}
\caption{Jet quenching parameter in the $T$-$\mu_B$ plane for a jet quark with initial momentum $|\bm{q}| = 3\GeV$: the left panel shows the $\pi$-meson contribution $\hat{q}_\pi$, and the right panel shows the $\sigma$-meson contribution $\hat{q}_\sigma$. The black solid line indicates the chiral phase boundary, and the red circle marks the CEP at $(T_\mathrm{CEP}, \mu_{B_\mathrm{CEP}}) = (98, 643)\MeV$.}
\label{fig:qhat}
\end{figure*}
%

In \Fig{fig:spectral} we show the $\sigma$ and $\pi$ spectral functions at $\mu_B = 636\MeV$ and external spatial momentum $|\bm{p}| = 0.3\GeV$ for three representative temperatures, with the $\mu_B$ value chosen as the closest sampled chemical potential to the CEP in our calculation, thereby providing a clear view of the spectral behavior in different thermodynamic regions near the CEP. Note that the CEP is located at $(T_\mathrm{CEP},\mu_{B_{\mathrm{CEP}}})=(98,643)$ MeV in the QCD-assisted low-energy effective theory \cite{Fu:2023lcm}. Since $\hat{q}$ is obtained as a weighted integral over the spectral function, these results also clarify the contribution of the spectral functions to the enhancement of $\hat{q}$. The left panels display the full spectral functions on a logarithmic scale, where the time-like peaks corresponding to the mesonic quasi-particle excitations are clearly visible at $\omega > |\bm{p}|$. The right panels zoom into the spacelike region $|\omega| \leq |\bm{p}|$ on a linear scale, which is the kinematic domain that contributes to $\hat{q}$ as discussed in \sec{sec:jet}.
 
We first discuss the time-like peaks of the $\sigma$ spectral function shown in \Fig{fig:spectral}~(a). At $T = 100\MeV$, which lies in the hadronic phase at this $\mu_B$, the $\sigma$ meson has a relatively large mass and the corresponding peak is located at higher energies. As the temperature increases to $T = 125\MeV$, close to the chiral phase boundary, the $\sigma$ peak shifts to significantly lower energies and becomes higher and narrower. This reflects the softening of the $\sigma$ mode as the system approaches the chiral phase transition, where the $\sigma$ meson plays the role of the critical mode and its mass is substantially reduced in the vicinity of the CEP. Upon further increasing the temperature to $T = 185\MeV$, well into the chirally restored regime, the $\sigma$ peak moves back to higher energies, broadens considerably, and decreases in height, indicating a larger thermal width at high temperature.

%
\begin{figure*}[t]
\centering
\includegraphics[width=1.0\textwidth]{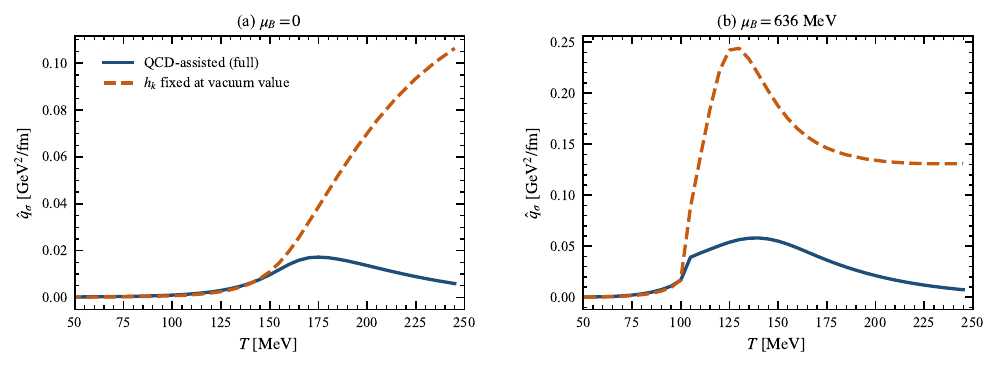}
\caption{Test of the role of the temperature-dependent Yukawa coupling $h_k(T)$ in shaping $\hat{q}_\sigma$, shown at (a) $\mu_B = 0$ and (b) $\mu_B = 636$~MeV, for a jet quark with initial momentum $|\bm{q}| = 3\GeV$. The blue solid curves show the full QCD-assisted calculation with $h_k(T)$ matched to $2+1$-flavor QCD, while the orange dashed curves show a diagnostic calculation in which $h_k$ is held fixed at its vacuum value $h_k(T{=}0)$.}
\label{fig:qhat_hfixed_compare}
\end{figure*}
%

The $\pi$ spectral function, shown in \Fig{fig:spectral}~(b), exhibits a qualitatively different behavior. At $T = 100\MeV$ the pion pole appears as a sharp and pronounced peak close to the light cone, reflecting the pseudo-Goldstone nature of the pion in the broken phase. As the temperature increases to $T = 125\MeV$, the pion peak shifts slightly to higher energies, reflecting the fact that the pion mode gradually loses its pseudo-Goldstone character and moves toward degeneracy with its chiral partner, the $\sigma$ mode, as the chiral symmetry is partially restored. Unlike the $\sigma$ meson, the pion does not exhibit a dramatic softening near the phase boundary, since it is not the critical mode of the chiral phase transition.
 
In the spacelike region, shown in the right panels of \Fig{fig:spectral}, the spectral functions at $T = 125\MeV$ (blue dashed) are significantly enhanced compared to those at $T = 100\MeV$ and $T = 185\MeV$. A notable exception is the pion spectral function at $T = 100\MeV$, where the spacelike part does not develop a clearly separated bump but instead connects smoothly to the timelike peak. The origin of this behavior can be understood from the structure of the mesonic loop functions $J_{\alpha\beta}^{B}$ (see Appendix~\ref{app:flow-2point}), whose imaginary part contains terms involving $\delta(E_{k,\alpha} \pm \omega \pm E_{p+q,\beta})$ and its derivatives. Combined with the theta functions from the $3d$ regulator, these delta functions contribute within the $\bm q$ integration domain only for $\omega$ in a certain range. For instance, for the term involving $\delta(E_{k,\alpha} + \omega - E_{p+q,\beta})$, this constrains $\omega$ to
\begin{widetext}
\begin{align}
    & \sqrt{(k-|\bm p|)^2 + \bar{m}^2_{k,\beta} } - \sqrt{k^2 + \bar{m}^2_{k,\alpha}} \;\leq\; \omega \;\leq\; \sqrt{(k+|\bm p|)^2 + \bar{m}^2_{k,\beta} } - \sqrt{k^2 + \bar{m}^2_{k,\alpha}}, \quad 0 < k \leq |\bm p|/2, \\[3pt]
    & \sqrt{k^2 + \bar{m}^2_{k,\beta}} - \sqrt{k^2 + \bar{m}^2_{k,\alpha}} \;\leq\; \omega \;\leq\; \sqrt{(k+|\bm p|)^2 + \bar{m}^2_{k,\beta} } - \sqrt{k^2 + \bar{m}^2_{k,\alpha}}, \quad k > |\bm p|/2.
\end{align}
\end{widetext}
When $\alpha = \beta$, the upper bound of $\omega$ is strictly less than $p$, so only the spacelike region is kinematically allowed; when $\alpha \neq \beta$, the upper bound can exceed $p$, i.e. the $\omega$ here may straddle the light cone, connecting the spacelike and timelike regions. As can be seen from the mesonic loop diagrams in \Fig{fig:Gamma2_flow}, the loops contributing to $\Gamma^{(2)}_\sigma$ involve only same-flavor internal lines ($\sigma\sigma$, $\pi\pi$), while those contributing to $\Gamma^{(2)}_\pi$ involve mixed flavors ($\sigma\pi$), so this cross-light-cone behavior appears only in $\rho_\pi$. As the temperature increases, $m_\sigma$ and $m_\pi$ approach each other, progressively suppressing this cross-light-cone effect and restoring a clearly separated spacelike bump. To further verify the consistency of our results, in \Fig{fig:degeneracy} we compare the $\sigma$ and $\pi$ spectral functions at $T = 185\MeV$. The two are nearly degenerate across the entire frequency range, as expected from the approximate restoration of chiral symmetry in the QGP phase. 
 
In \Fig{fig:qhat} we present the $\sigma$- and $\pi$-meson contributions to the jet quenching parameter in the $T$-$\mu_B$ plane for a jet quark with initial momentum $|\bm{q}| = 3\GeV$, following Ref.~\cite{Wu:2022vbu}. As can be seen, $\hat{q}$ takes appreciable values mainly above the chiral phase boundary, indicating that energetic jets primarily interact with the deconfined QGP medium and are much less sensitive to hadronic matter. In the chirally restored phase, both $\hat{q}_\sigma$ and $\hat{q}_\pi$ exhibit a pronounced enhancement at large baryon chemical potential as the chiral crossover sharpens toward the CEP. Within the range of $\mu_B$ accessible to the present calculation ($\mu_B \lesssim 636$~MeV), our results are consistent with the PCO scenario. However, a definitive identification of the genuine CEP-driven critical enhancement, as distinct from the gradual enhancement already produced by the sharpening crossover below the CEP, requires extending the calculation to the immediate vicinity of and beyond the CEP, which is for the moment challenging in the numerical calculations and will be reported elsewhere in the future.

The high-temperature behavior of $\hat{q}$ in Fig.~\ref{fig:qhat} deserves a brief comment. We find that $\hat{q}$ decreases at high $T$, a feature that originates from the QCD-running Yukawa coupling $h_k^{QCD}(T)$ used in the QCD-assisted low energy effective theory. The $h_k^{QCD}(T)$ is obtained from QCD calculations in \cite{Fu:2019hdw}. Thus, the QCD-assisted low energy effective theory incorporates an important feature of QCD that is absent in a pure quark-meson treatment: the gradual weakening of the quark-meson coupling at high temperatures. Specifically, this $h_k^{QCD}(T)$ input enters $\hat{q}$ at two levels. First, it modifies the internal structure of the fRG flow---the effective potential, the wave-function renormalizations, and the running quark mass. Second, $h_k(T)$ appears explicitly in the final expression for $\hat{q}$, both through the overall coupling $g_{\text{eff}}^2 = \bar{h}_{k=0}^2/4$ and through its explicit appearance in the spectral functions.

To identify which of these two roles of $h_k(T)$ is responsible for the enhancement of $\hat{q}$ at finite $\mu_B$, we perform a test, as can be seen in \Fig{fig:qhat_hfixed_compare}: we keep all results of the full QCD-assisted flow unchanged, but fix $h_k$ to its vacuum value---removing its temperature dependence---in the final $\hat{q}$ expression and in the spectral functions used therein. This isolates the direct effect of the QCD-running coupling on $\hat{q}$ while preserving its imprint on the internal flow structure. At $\mu_B = 0$, the $h_k$ fixed calculation yields a $\hat{q}$ that grows monotonically with temperature throughout the chirally restored phase, in contrast to the high-temperature suppression seen in the full calculation; at this vanishing chemical potential, once the temperature dependence of $h_k(T)$ is removed, no peak near the phase boundary remains. At $\mu_B = 636$~MeV, in contrast, the $h_k$ fixed  calculation exhibits a pronounced peak near the chiral crossover followed by a plateau-like behavior at high temperatures, which is different from the situation at $\mu_B = 0$. This demonstrates that the enhancement of $\hat{q}$ at large $\mu_B$ originates from the QCD-induced modifications to the internal flow structure---rather than from the explicit temperature dependence of $h_k(T)$ in the final expression.

A comparison of the two panels reveals that $\hat{q}_\pi$ is approximately three times as large as $\hat{q}_\sigma$ across the phase diagram, with both exhibiting a similar pattern of critical enhancement near the CEP. The ratio is consistent with the pion degeneracy factor $d_\pi = 3$, indicating that the per-component contributions from the two channels are comparable in magnitude. This can be understood as follows. Although the $\sigma$ meson is the critical mode whose mass vanishes at the CEP, the pion is not an independent spectator: as shown in the flow equation for the pion two-point function in \app{app:flow-2point}, the $\sigma$ propagator $G_\sigma$ appears in the internal lines of the bosonic loop diagrams $J^B_{\sigma\pi}$, $J^B_{\pi\sigma}$ and the tadpole term $I^{(2)}_\sigma$. Near the CEP, the $\sigma$ propagator $G_\sigma \sim 1/m^2_\sigma$ grows rapidly as the $\sigma$ mass softens, which amplifies these contributions and thereby enhances the pion spectral function. In other words, the critical fluctuations of the order parameter $\sigma$ are transmitted to the pion channel through their coupling as components of the chiral multiplet $\phi = (\sigma, \bm{\pi})$.
 
Moreover, the enhancement is not confined to the immediate vicinity of the CEP but extends along the phase boundary towards smaller $\mu_B$. This is driven by the successive sharpening of the chiral crossover with increasing $\mu_B$, which leads to a sizable enhancement of $\hat{q}$ even far from the CEP. This means that the signal of enhanced jet quenching is not restricted to a narrow window of collision energies but could be observable over a broader range of the beam energy scan.

It is worth noting that in the mean-field calculation of Ref.~\cite{Wu:2022vbu}, no enhancement of $\hat{q}$ near the CEP was observed without coupling the quark-meson model to the Ising model. In contrast, the fRG approach employed in this work systematically incorporates quantum and thermal fluctuations through the Wetterich flow equation, which naturally generates the critical behavior without resorting to an external universality-class mapping. As a result, the enhancement 
of $\hat{q}$ near the CEP emerges directly from the fRG calculation itself.
 
\section{Summary and outlook}
\label{sec:summary}

We have computed the jet quenching parameter $\hat{q}$ at finite temperature and baryon chemical potential within a QCD-assisted PQM model. Unlike the previous mean-field treatment in Ref.~\cite{Wu:2022vbu}, which required an external mapping to the Ising model to capture critical behavior, the fRG approach employed here generates the enhancement of $\hat{q}$ near the CEP directly through the Wetterich flow equation.

The resulting $\hat{q}$ takes appreciable values mainly above the chiral phase boundary, indicating that energetic jets primarily interact with the deconfined QGP medium, and both the $\sigma$- and $\pi$-meson contributions exhibit a pronounced enhancement at large baryon chemical potential as the chiral crossover sharpens toward the CEP. Notably, this enhancement extends along the phase boundary toward smaller $\mu_B$, which suggests that the signal of enhanced jet quenching could be accessible over a broader range of beam energies, rather than being confined to a narrow window near the CEP.

Several extensions of this work are warranted. Most importantly, the present calculation includes only mesonic contributions to $\hat{q}$ from quark--meson scattering; it is necessary for a quantitative comparison with phenomenologically extracted values to incorporate quark-gluon scattering channels~\cite{Majumder:2010qh, Burke:2013yra}. Moreover, connecting to the nuclear modification factor $R_{AA}$ will require realistic hydrodynamic modeling and a careful treatment of initial-state effects, in particular the Cronin enhancement~\cite{Cronin:1973fd, Antreasyan:1978cw, Accardi:2003jh, Vitev:2003xu}, whose interplay with the critical enhancement of $\hat{q}$ at BES energies deserves a dedicated study~\cite{Ke:2022gkq, Vitev:2005he}.

\section*{Acknowledgements}

We thank Yong-rui Chen and Zi-ning Wang for discussions. This work is supported by the National Natural Science Foundation of China under Grant No.\ 12447102.\\


\appendix

\section{Flow equations of two-point correlation functions}
\label{app:flow-2point}

The mathematical representation of the mesonic 2-point correlation functions in \Fig{fig:Gamma2_flow} is given by
\begin{align}
\frac{\partial_t \Gamma_{\sigma,k}^{(2)}}{Z_{\phi,k}}
&= -N_c N_f J_{\sigma}^{F}(p)- \frac{1}{2} I_{\sigma}^{(2)} \Gamma_{\sigma\sigma\sigma\sigma}^{(4)}
       - \frac{3}{2} I_{\pi}^{(2)} \Gamma_{\sigma\sigma\pi\pi}^{(4)} \nonumber\\
&\quad + J_{\sigma\sigma}^{B}(p)\big(\Gamma_{\sigma\sigma\sigma}^{(3)}\big)^2
   + 3 J_{\pi\pi}^{B}(p)\big(\Gamma_{\sigma\pi\pi}^{(3)}\big)^2 ,\\
\frac{\partial_t \Gamma_{\pi,k}^{(2)}}{Z_{\phi,k}}
&= -N_c N_f J_{\pi}^{F}(p)
       - \frac{1}{2} I_{\pi}^{(2)}\Big( 2\,\Gamma_{\pi\pi\tilde{\pi}\tilde{\pi}}^{(4)} + \Gamma_{\pi\pi\pi\pi}^{(4)}\Big) \nonumber\\
&\quad - \frac{1}{2} I_{\sigma}^{(2)}\Gamma_{\sigma\sigma\pi\pi}^{(4)}
+ \big(J_{\sigma\pi}^{B}(p) + J_{\pi\sigma}^{B}(p)\big)\big(\Gamma_{\pi\pi\sigma}^{(3)}\big)^2 ,
\end{align}
where $\pi,\tilde{\pi}\in\{\pi_1,\pi_2,\pi_3\}$ with $\pi\neq\tilde{\pi}$, and the loop functions $J_{\alpha}^{F}(p),J_{\alpha\beta}^{B}(p),I_{\gamma}^{(i)}$ for $\alpha,\beta\in\{\sigma,\pi\}$ and $\gamma\in\{\sigma,\pi,q\}$ are defined as
\begin{align}
J_{\alpha\beta}^{B}(p)
  &= \mathrm{Tr}_{q}\!\Big[\partial_t R_k^{B}(q)\,
     G_{\alpha,k}(q\!+\!p)\,G_{\beta,k}^{2}(q)\Big],
     \label{eq:JB} \\[4pt]
J_{\alpha}^{F}(p)
  &= \mathrm{Tr}_{q}\!\Big[\partial_t R_k^{F}(q)\,
     G_{q,k}(q)\,\Gamma_{\bar{q}q\alpha}^{(3)}
     \nonumber\\
  &\qquad\times
     G_{q,k}(q\!+\!p)\,
     \Gamma_{\bar{q}q\alpha}^{(3)}\,
     G_{q,k}(q)\Big]
     \nonumber\\
  &\quad + (p\to -p),
     \label{eq:JF} \\[4pt]
I_{\gamma}^{(i)}
  &= \mathrm{Tr}_{q}\!\Big[\partial_t R_k^{A}(q)\,
     G_{\gamma,k}^{i}(q)\Big],
     \label{eq:In}
\end{align}
where $G_{\gamma,k}(q)=(\Gamma_{\gamma,k}^{(2)}(q)+R_k^{A}(q))^{-1}$ is the regulated propagator and $A\in\{B,F\}$ is determined by whether the field $\gamma$ is bosonic or fermionic. In this work, we use the $3d$ regulators as follows,
\begin{align}
    R_k^B(q)&=Z_{\phi,k}\,\bm{q}^2 \,r_{B}(\bm{q}^2/k^2)\,, \label{eq:RkB}\\[2ex]
    R_k^F(q)&=Z_{q,k}\,\mathrm{i} \bm{\gamma} \cdot \bm{q} \,r_{F}(\bm{q}^2/k^2)\,,\label{eq:RkF}
\end{align}
with
\begin{align}
    r_B(x)&=(\frac{1}{x}-1)\theta(1-x)\,, \\[2ex]
    r_F(x)&=(\frac{1}{\sqrt{x}}-1)\theta(1-x)\,. 
\end{align}
Before presenting the explicit expressions for the loop functions, we give 
the definitions of the renormalized couplings, masses, mesonic vertices, 
and anomalous dimensions used throughout this work, which is convenient 
for expressing the loop functions in a compact form.
The renormalized Yukawa coupling reads
\begin{align}
    \bar{h}_k = \frac{h_k}{Z_{q,k}\,Z_{\phi,k}^{1/2}}\,,
    \label{eq:hbar}
\end{align}
and the renormalized meson and quark masses are given by
\begin{align}
    \bar{m}^2_{k,\pi} &= \frac{V_k'(\kappa_k)}{\,Z_{\phi,k}}\,,\nonumber\\[2pt]
    \bar{m}^2_{k,\sigma} &= \frac{V_k'(\kappa_k) + 2\kappa_k\,V_k''(\kappa_k)}{\,Z_{\phi,k}}\,,\nonumber\\[0pt]
    \bar{m}^2_{k,q} &= \frac{h_k^2\,\kappa_k}{2\,Z_{q,k}^2}\,,
    \label{eq:masses}
\end{align}
where $\kappa_k$ is the scale-dependent physical minimum of $V_k(\rho)$.
The three- and four-point mesonic vertex functions appearing in the flow equations of the two-point functions are obtained from the derivatives of the effective potential. Their explicit expressions read
\begin{align}
    \Gamma^{(3)}_{\sigma\sigma\sigma} &= Z_{\phi,k}^{-3/2}\,\big(3V_k''(\kappa_k)\,\sigma_0 + V_k^{(3)}(\kappa_k)\,\sigma_0^3\big)\,,\nonumber\\[3pt]
    \Gamma^{(3)}_{\sigma\pi\pi} &= Z_{\phi,k}^{-3/2}\,V_k''(\kappa_k)\,\sigma_0\,,\nonumber\\[3pt]
    \Gamma^{(4)}_{\sigma\sigma\sigma\sigma} &= Z_{\phi,k}^{-2}\,\big(3V_k''(\kappa_k) + 6V_k^{(3)}(\kappa_k)\,\sigma_0^2 + V_k^{(4)}(\kappa_k)\,\sigma_0^4\big)\,,\nonumber\\[3pt]
    \Gamma^{(4)}_{\sigma\sigma\pi\pi} &= Z_{\phi,k}^{-2}\,\big(V_k''(\kappa_k) + V_k^{(3)}(\kappa_k)\,\sigma_0^2\big)\,,\nonumber\\[3pt]
    \Gamma^{(4)}_{\pi\pi\tilde{\pi}\tilde{\pi}} &= Z_{\phi,k}^{-2}\,V_k''(\kappa_k)\,,\nonumber\\[3pt]
    \Gamma^{(4)}_{\pi\pi\pi\pi} &=
    3\,Z_{\phi,k}^{-2}\,V_k''(\kappa_k)\,,
    \label{eq:vertices_combined}
\end{align}
where $\sigma_0 = \sqrt{2\kappa_k}$ is the expectation value of the $\sigma$ field evaluated at the minimum of the effective potential. In this work, we emphasize that all derivatives of $V_k$ are taken with respect to $\rho$, rather than the $\sigma$ field.

The anomalous dimensions for the meson and quark fields are defined as the logarithmic scale derivatives of the respective wave functions,
\begin{align}
    \eta_{\phi,k} = -\frac{\partial_t Z_{\phi,k}}{Z_{\phi,k}}\,,\qquad
    \eta_{q,k} = -\frac{\partial_t Z_{q,k}}{Z_{q,k}}\,.
    \label{eq:anomdim}
\end{align}
In the following, we present the explicit expressions for the loop functions.
\newcommand{\rfac}{\widetilde{\eta}_{q,k}}
\newcommand{\rbac}{\widetilde{\eta}_{\phi,k}}
\begin{align}
J_{\sigma}^{F}(p)
  &= 4k^{2}\bar{h}_k^{\,2}
     \!\int_{D_1}\!\frac{d^{3}\bm{q}}{(2\pi)^{3}}\,
     \rfac\;
     \Big[\,\ell^{(2,0)}(E_{k,q})
     \nonumber\\
  &\quad
     + (1\!-\!F_1)\,\ell^{(1,1)}(E_{k,q},E_{k,q})+ \big(2E_{k,q}^{2} \nonumber\\
  &\quad+ 2\bar{m}_{k,q}^{2}
       + p_0^{2} - 2k^{2}F_1\big)
     \ell^{(2,1)}(E_{k,q},E_{k,q})\,\Big]
     \nonumber\\
   &\quad
   +4k\,\bar{h}_k^{\,2}
     \!\int_{D_2}\!\frac{d^{3}\bm{q}}{(2\pi)^{3}}\,
     \rfac\;
     \Big[\,k\,\ell^{(2,0)}(E_{k,q})
     \nonumber\\
  &\quad
     + (k\!-\!F_2)\,
       \ell^{(2,1)}(E_{k,q},E_{p+q,q})
     \nonumber\\
  &\quad
     + \big((E_{k,q}^{2}\!+\!E_{p+q,q}^{2}
       \!+\!2\bar{m}_{k,q}^{2}\!+\!p_0^{2})k
     \nonumber\\
  &\qquad
     - 2k^{2}F_2\big)\,
     \ell^{(1,1)}(E_{k,q},E_{p+q,q})
     \,\Big],
   \\[10pt]
J_{\pi}^{F}(p)
  &= 4k^{2}\bar{h}_k^{\,2}
     \!\int_{D_1}\!\frac{d^{3}\bm{q}}{(2\pi)^{3}}\,
     \rfac\;
     \Big[\,\ell^{(2,0)}(E_{k,q})
     \nonumber\\
  &\quad
     + (1\!-\!F_1)\,\ell^{(1,1)}(E_{k,q},E_{k,q})
     \nonumber\\
  &\quad
     + \big(2E_{k,q}^{2} + 2\bar{m}_{k,q}^{2}
       + p_0^{2} - 2(E_{k,q}^{2}+\bar{m}_{k,q}^{2})F_1\big)
       \nonumber\\
  &\quad\times
     \ell^{(2,1)}(E_{k,q},E_{k,q})\,\Big]
     \nonumber\\
   &\quad
   +4k\,\bar{h}_k^{\,2}
     \!\int_{D_2}\!\frac{d^{3}\bm{q}}{(2\pi)^{3}}\,
     \rfac\;
     \Big[\,k\,\ell^{(2,0)}(E_{k,q})
     \nonumber\\
  &\quad
     + (k\!-\!F_2)\,
       \ell^{(2,1)}(E_{k,q},E_{p+q,q})
     \nonumber\\
  &\quad
     + \big((E_{k,q}^{2}\!+\!E_{p+q,q}^{2}
       \!-\!2\bar{m}_{k,q}^{2}\!+\!p_0^{2})k
     \nonumber\\
  &\qquad
     - 2k^{2}F_2\big)\,
     \ell^{(1,1)}(E_{k,q},E_{p+q,q})
     \,\Big], 
     \\[10pt]
J_{\alpha\beta}^{B}(p)
  &=k^2
     \!\int_{D_1}\!\frac{d^{3}\bm{q}}{(2\pi)^{3}}\,
     \rbac\;
     (\tilde{\ell}^{(2,1)}(E_{k,\beta},E_{k,\alpha}))
   \nonumber\\
  &\quad
  +k^2
  \!\int_{D_2}\!\frac{d^{3}\bm{q}}{(2\pi)^{3}}\,
     \rbac\;
     (\tilde{\ell}^{(2,1)}(E_{k,\beta},E_{p+q,\alpha})), 
     \\[10pt]
I_{\alpha}^{(2)}
  &=-k^2
     \!\int_{D_3}\!\frac{d^{3}\bm{q}}{(2\pi)^{3}}\,
     \rbac\;
     (\tilde{\ell}^{(2,0)}(E_{k,\alpha})).   
\end{align}
Here we define the factors $\rfac$ and $\rbac$ as
\begin{align}
    \rfac &\equiv 1 - \eta_{q,k} + \frac{|\bm{q}|}{k}\,\eta_{q,k}\,,\\[2ex]
    \rbac &\equiv 2 - \eta_{\phi,k} - \frac{\bm{q}^2}{k^2}\,\eta_{\phi,k}\,.
\end{align}
The integration domains are defined by
\begin{align}
    D_1: &\quad\theta(k^{2}-(\bm{p}+\bm{q})^{2})\theta(k^{2}-\bm{q}^{2})\,,\\[2ex]
    D_2: &\quad\theta(-{k}^{2}+(\bm{p}+\bm{q})^{2})\theta(k^{2}-\bm{q}^{2})\,,\\[2ex]
    D_3: &\quad\theta(k^{2}-\bm{q}^{2})\,,
\end{align}
respectively. The ${F}_{1}$ and ${F}_{2}$ are defined as
\begin{align} 
 {F}_{1}\equiv\frac{(\bm{p}+\bm{q})\cdot\bm{q}}{|\bm{p}+\bm{q}||\bm{q}|}\,,\qquad {F}_{2}\equiv\frac{(\bm{p}+\bm{q})\cdot\bm{q}}{|\bm{q}|}\,.
\end{align}
The effective quasi-particle energies $E_{k,\alpha}$ and $E_{p+q,\alpha}$ with $\alpha\in\{\sigma,\pi,q\}$ are defined as
\begin{align}
E_{k,\alpha}&\equiv\sqrt{k^2+\bar{m}^2_{k,\alpha}}\,,\\[2ex]
E_{p+q,\alpha}&\equiv\sqrt{(\bm{p}+\bm{q})^2+\bar{m}^2_{k,\alpha}}\,.
\end{align}
The fermionic threshold functions $\ell$ in the loop functions are given by
\begin{widetext}
\begin{align}
    \ell^{(i,j)}(E_1,E_2) &\equiv -T \sum_{n \in \mathbb{Z}} \frac{1}{\big( (i q_0 - \mu)^2 - E_1^2 \big)^i \big( (i q_0 + i p_0 - \mu)^2 - E_2^2  \big)^j }+(p_0\to -p_0),
\end{align}
and the bosonic one $\tilde{\ell}^{(i,j)}(E_1,E_2) = \tfrac{1}{2}\, \ell^{(i,j)}(E_1,E_2)\big|_{\mu=0}$ \,, with $q_0 = (2n+1)\pi T$ in $\ell$ and $2n\pi T$ in  $\tilde{\ell}$\,($n\in\mathbb{Z}$). In the following, we present the explicit expressions for the threshold functions after performing the Matsubara summation.
{\setlength{\jot}{4pt}%

\begin{flalign}
&\ell^{(2,0)}(E_{k,q})
  = -\frac{1}{2E_{k,q}^{3}}
     \big(1 - n_F(E_{k,q}+\mu) - n_F(E_{k,q}-\mu)\big)
     -\frac{1}{2E_{k,q}^{2}}\big(n_F'(E_{k,q}+\mu)
     + n_F'(E_{k,q}-\mu)\big)\,,
     &&\label{eq:ell20}
\\[10pt]
&\ell^{(1,1)}(E_{k,q},E_{k,q})
  = -\frac{2}{4E_{k,q}^{3}+E_{k,q}\,p_0^{2}}
     \big(1 - n_F(E_{k,q}+\mu)
     - n_F(E_{k,q}-\mu)\big)\,,
     &&\label{eq:ell11}
\\[10pt]
&\ell^{(2,1)}(E_{k,q},E_{k,q})
  = \frac{12E_{k,q}^{2}+p_0^{2}}
     {2E_{k,q}^{3}(4E_{k,q}^{2}+p_0^{2})^{2}}
     \big(1-n_F(E_{k,q}+\mu)
     -n_F(E_{k,q}-\mu)\big)
     &&\notag\\
&\hspace{7em}
     +\frac{1}{2E_{k,q}^{2}(4E_{k,q}^{2}+p_0^{2})}
     \big(n_F'(E_{k,q}+\mu)
     +n_F'(E_{k,q}-\mu)\big)\,,
     &&\label{eq:ell21}
\\[10pt]
&\ell^{(1,1)}(E_{k,q},E_{p+q,q}) 
  = -\frac{1-n_F(E_{k,q}+\mu)-n_F(E_{k,q}-\mu)}{2E_{k,q}}
     \left[\frac{1}{E_{p+q,q}^2-(E_{k,q}-ip_0)^2}
     +\frac{1}{E_{p+q,q}^2-(E_{k,q}+ip_0)^2}\right]
     &&\notag\\
&\hspace{5em}
     -\frac{1-n_F(E_{p+q,q}+\mu)-n_F(E_{p+q,q}-\mu)}{2E_{p+q,q}}
     \left[\frac{1}{E_{k,q}^2-(E_{p+q,q}+ip_0)^2}
     +\frac{1}{E_{k,q}^2-(E_{p+q,q}-ip_0)^2}\right]\,,
     &&\label{eq:ell11pq}
\\[10pt]
&\ell^{(2,1)}(E_{k,q},E_{p+q,q})
  = \frac{1-n_F(E_{p+q,q}+\mu)-n_F(E_{p+q,q}-\mu)}{2E_{p+q,q}}
     \left[\frac{1}{\big(E_{k,q}^2-(E_{p+q,q}-ip_0)^2\big)^2}
     +\frac{1}{\big(E_{k,q}^2-(E_{p+q,q}+ip_0)^2\big)^2}\right]
     &&\notag\\[8pt]
&\quad
     +\frac{1-n_F(E_{k,q}+\mu)-n_F(E_{k,q}-\mu)}{4E_{k,q}^3}
     \left[\frac{E_{p+q,q}^2-3E_{k,q}^2-4iE_{k,q}p_0+p_0^2}
     {\big(E_{p+q,q}^2-(E_{k,q}+ip_0)^2\big)^2}
     +\frac{E_{p+q,q}^2-3E_{k,q}^2+4iE_{k,q}p_0+p_0^2}
     {\big(E_{p+q,q}^2-(E_{k,q}-ip_0)^2\big)^2}\right]
     &&\notag\\[8pt]
&\quad
     +\frac{n_F'(E_{k,q}+\mu)+n_F'(E_{k,q}-\mu)}{4E_{k,q}^2}
     \left[\frac{1}{E_{p+q,q}^2-(E_{k,q}+ip_0)^2}
     +\frac{1}{E_{p+q,q}^2-(E_{k,q}-ip_0)^2}\right]\,,
\\[10pt]
&\tilde{\ell}^{(2,1)}(E_{k,\beta},E_{p+q,\alpha})
  = \frac{1+2n_B(E_{p+q,\alpha})}{4E_{p+q,\alpha}}
    \left[
    \frac{1}{\big(E_{k,\beta}^2-(E_{p+q,\alpha}-ip_0)^2\big)^2}
    +\frac{1}{\big(E_{k,\beta}^2-(E_{p+q,\alpha}+ip_0)^2\big)^2}
    \right]
    &&\notag\\[8pt]
&\hspace{8.25em}
    +\frac{1+2n_B(E_{k,\beta})}{8E_{k,\beta}^3}
    \left[
    \frac{E_{p+q,\alpha}^2-3E_{k,\beta}^2-4iE_{k,\beta}p_0+p_0^2}
    {\big(E_{p+q,\alpha}^2-(E_{k,\beta}+ip_0)^2\big)^2}
    +\frac{E_{p+q,\alpha}^2-3E_{k,\beta}^2+4iE_{k,\beta}p_0+p_0^2}
    {\big(E_{p+q,\alpha}^2-(E_{k,\beta}-ip_0)^2\big)^2}
    \right]
    &&\notag\\[4pt]
&\hspace{8.15em}
    -\frac{n_B'(E_{k,\beta})}{4E_{k,\beta}^2}
    \left[
    \frac{1}{E_{p+q,\alpha}^2-(E_{k,\beta}+ip_0)^2}
    +\frac{1}{E_{p+q,\alpha}^2-(E_{k,\beta}-ip_0)^2}
    \right]\,,
\\[10pt]
&\tilde{\ell}^{(2,1)}(E_{k,\beta},E_{k,\alpha})
  = \frac{1+2n_B(E_{k,\alpha})}{4E_{k,\alpha}}
    \left[
    \frac{1}{\big(E_{k,\beta}^2-(E_{k,\alpha}-ip_0)^2\big)^2}
    +\frac{1}{\big(E_{k,\beta}^2-(E_{k,\alpha}+ip_0)^2\big)^2}
    \right]
    &&\notag\\[8pt]
&\hspace{7.25em}
    +\frac{1+2n_B(E_{k,\beta})}{8E_{k,\beta}^3}
    \left[
    \frac{E_{k,\alpha}^2-3E_{k,\beta}^2-4iE_{k,\beta}p_0+p_0^2}
    {\big(E_{k,\alpha}^2-(E_{k,\beta}+ip_0)^2\big)^2}
    +\frac{E_{k,\alpha}^2-3E_{k,\beta}^2+4iE_{k,\beta}p_0+p_0^2}
    {\big(E_{k,\alpha}^2-(E_{k,\beta}-ip_0)^2\big)^2}
    \right]
    &&\notag\\[4pt]
&\hspace{7.15em}
    -\frac{n_B'(E_{k,\beta})}{4E_{k,\beta}^2}
    \left[
    \frac{1}{E_{k,\alpha}^2-(E_{k,\beta}+ip_0)^2}
    +\frac{1}{E_{k,\alpha}^2-(E_{k,\beta}-ip_0)^2}
    \right]\,,
\\[10pt]
&\tilde{\ell}^{(2,0)}(E_{k,\alpha})
  = -\frac{1+2n_B(E_{k,\alpha})}{4E_{k,\alpha}^3}
    +\frac{n_B'(E_{k,\alpha})}{2E_{k,\alpha}^2}\,.
\end{flalign}}
\end{widetext}
where $\alpha,\beta\in\{\sigma,\pi\}$. The bosonic and
fermionic distribution functions are given by
\begin{align}
    n_B(x)=\frac{1}{e^{x/T}-1}\,,
\end{align}
and
\begin{align}
n_F(x)=
\begin{cases}
    \frac{1 + 2\bar{L}\,e^{x/T}+ L\,e^{2x/T}}{1 + 3\bar{L}\,e^{x/T}+ 3L\,e^{2x/T} + e^{3x/T}},\,(x=E-\mu)\\[2ex]
    \frac{1 + 2 L\,e^{x/T}+ \bar{L}\,e^{2x/T}}{1 + 3L\,e^{x/T}+ 3\bar{L}\,e^{2x/T} + e^{3x/T}},\,(x=E+\mu)  
\end{cases}\,,
\end{align}
where $L$ and $\bar{L}$ denote the Polyakov loop
and its conjugate, which arise from taking the
color trace over the temporal gluon background
field $A_0$ in the PQM effective action
Eq.~(\ref{eq:action}). Their explicit definitions
are given in Appendix~\ref{app:setup}.

\section{Details of the theoretical setup}
\label{app:setup}

Our theoretical setup follows that of Ref.~\cite{Fu:2023lcm}, 
and we summarize the key ingredients here. In the PQM model, the coupling between the quark sector and the gluon sector is realized through the Polyakov loop and its conjugate. The traced Polyakov loop $L$ and its conjugate $\bar{L}$ are defined as
\begin{align}
    L(\bm{x}) = \frac{1}{N_c} \langle \mathrm{Tr}_c \, \mathcal{P}(\bm{x}) \rangle\,, \,\,
    \bar{L}(\bm{x}) = \frac{1}{N_c} \langle \mathrm{Tr}_c \, \mathcal{P}^\dagger(\bm{x}) \rangle\,,
    \label{eq:polyakov_loop_def}
\end{align}
where $\mathrm{Tr}_c$ denotes the trace in color space, and the Polyakov loop matrix $\mathcal{P}(\bm{x})$ is given by
\begin{align}
    \mathcal{P}(\bm{x}) = \mathcal{P} \exp\left( ig \int_0^\beta d\tau \, A_0(\tau, \bm{x}) \right)\,,
    \label{eq:polyakov_loop_matrix}
\end{align}
with $\beta = 1/T$ being the inverse temperature, $g$ the gauge coupling, $A_0$ the temporal component of the gluon field, and $\mathcal{P}$ the path-ordering operator.

The scale- and temperature-dependent Yukawa coupling 
is determined by matching to first-principles $2+1$-flavor 
QCD~\cite{Fu:2019hdw},
\begin{align}
    h_k(T) = h_0 \, \frac{h_{k}^{\mathrm{QCD}}(T)}
    {h_{0}^{\mathrm{QCD}}(0)}\,.
    \label{eq:hk_QCD}
\end{align}
The flow equations are integrated from an initial UV scale 
$\Lambda = 700\MeV$ with the initial effective potential 
\begin{align}
    V_{\mathrm{mat},\Lambda}(\rho) = \nu_\Lambda\,\rho 
    + \frac{\lambda_\Lambda}{2}\,\rho^2\,.
    \label{eq:V_init}
\end{align}

Together with the vacuum value of the Yukawa coupling $h_0$ in 
\Eq{eq:hk_QCD} and the strength of the explicit chiral symmetry 
breaking $c$ in \Eq{eq:action}, the parameters $\lambda_\Lambda$ 
and $\nu_\Lambda$ constitute the full set of input parameters, 
which are listed in \Tab{tab:parameters}. 

\begin{table}[H]
\centering
\begin{tabular*}{\columnwidth}{@{\hspace{1em}}c@{\hspace{3.2em}}c%
@{\hspace{2.5em}}c@{\hspace{3.2em}}c@{\hspace{1em}}}
\toprule
$\lambda_\Lambda$ & $\nu_\Lambda\;[\mathrm{GeV}^2]$ 
& $c\;[\times 10^{-3}\,\mathrm{GeV}^3]$ & $h_0$ \\
\midrule
10.15 & 0.53 & 1.6 & 11.6 \\
\bottomrule
\end{tabular*}
\caption{Input parameters in this work.}
\label{tab:parameters}
\end{table}

With the above input parameters, we obtain the physical 
observables in the vacuum as well as the location of the CEP 
in the phase diagram. The results are summarized in 
\Tab{tab:observables}.

%
\begin{table}[H]
\centering
\begin{tabular*}{\columnwidth}{@{\hspace{1em}\extracolsep{\fill}}%
ccccc@{\hspace{1em}}}
\toprule
$m_\pi$ & $m_\sigma$ & $(T_\mathrm{CEP},\,\mu_{B_\mathrm{CEP}})$ 
& $m_q$ & $\langle\sigma\rangle$ \\
\midrule
137 & 431 & (98,\,643) & 343 & 76 \\
\bottomrule
\end{tabular*}
\caption{Vacuum observables and the location of the CEP. All 
quantities are in units of MeV.}
\label{tab:observables}
\end{table}
%

\FloatBarrier

\bibliography{ref-lib}

\end{document}